# Continuous modal spectrum in nonreciprocal cavities

FILIPA R. PRUDÊNCIO,[1,2*] DAVID E. FERNANDES,[1] MÁRIO G. SILVEIRINHA,[1]

[1] *University of Lisbon – Instituto Superior Técnico and Instituto de Telecomunicações, Avenida Rovisco Pais 1, 1049-001 Lisbon, Portugal*

[2] *Instituto Universitário de Lisboa (ISCTE-IUL), Avenida das Forças Armadas 376, 1600-077 Lisbon, Portugal*

*filipa.prudencio@lx.it.pt

## Abstract

Nonreciprocal systems enable asymmetric energy transport and suppress backscattering, giving rise to unconventional wave phenomena. Here, we show that nonreciprocal cavities based on unidirectional waveguides exhibit a continuous modal spectrum, in contrast to conventional cavities with discrete eigenmodes. Using a ferrite-loaded microwave cavity as an example, we demonstrate that enforcing unidirectionality, by tailoring the waveguide geometry, drives a transition from discrete to continuous spectra, accompanied by strong spatial localization of electromagnetic fields. Our results reveal that dissipation alone fails to regularize these singular responses, highlighting the need for additional mechanisms to control localization in nonreciprocal systems.

## 1. Introduction

Unidirectional wave propagation in nonreciprocal photonic systems is one of the most impressive manifestations of nonreciprocity. Initially demonstrated in magnetically biased ferrite waveguides in the 1950s [1-4], and subsequently revisited within the framework of topological photonics [5–8], this effect underpins a wide variety of microwave and photonic devices, including electromagnetic isolators and circulators.

An important consequence of nonreciprocity is the emergence of singular wave responses associated with strong spatial localization. Previous studies [9-20] have shown that combining one-way waveguides with dissipative mechanisms can lead to the formation of electromagnetic "hotspots", where energy accumulates in highly confined regions of space. This effect is closely connected to broader discussions of nonreciprocal cavities, localization, and time-bandwidth constraints [21-22]. More recently, it has been recognized that these hotspots are closely linked to real-space singularities, which can be interpreted as manifestations of ill-defined topologies [17, 19, 20, 23]. These findings point to a deep connection between nonreciprocity, localization, and the emergence of singular field distributions. Furthermore, it was shown that, although dissipation ensures a well-defined electromagnetic response with square-integrable fields in externally driven one-way systems, the normal modes may not be square-integrable [23]. This behavior was linked to a thermodynamic paradox and to a breakdown of the conventional fluctuation-dissipation-theorem (FDT) description based on the idealized local material model [23].

In this work, we revisit this problem from a different perspective, focusing on how real-space singularities influence the modal structure of electromagnetic cavities. Contrary to the conventional paradigm, in which closed cavities support a discrete set of eigenmodes, we show that nonreciprocal cavities can exhibit a fundamentally different spectral behavior. Specifically, a cavity formed by a unidirectional waveguide terminated by reactive boundaries does not possess the usual discrete cavity spectrum. Instead, due to the absence of backscattering, the modal spectrum becomes continuous.

To illustrate this phenomenon, we analyze a microwave cavity partially filled with a ferrite material and investigate two limiting processes that enforce unidirectional propagation along the waveguide. One limiting process consists of initially introducing an additional slab of air inside the waveguide, whose thickness is subsequently reduced to zero. The second process is based on nonlocal corrections to the ferrite's response, with the unidirectional regime recovered in the limit where the nonlocal description converges to the corresponding local-response model. As the system approaches the unidirectional limit, we show that the spectrum undergoes a transition from discrete to continuous. In parallel, the associated electromagnetic fields become increasingly localized, with the magnetic field exhibiting a divergent behavior.

Our findings are consistent with Ref. [23], which showed that material dissipation alone is insufficient to ensure the square integrability of the cavity normal modes in the unidirectional limit, even when the driven response does not exhibit any obvious pathologies. We argue here that this property is a consequence of the cavity continuous modal spectrum. Our results thus provide additional insight into the interplay between nonreciprocity, spectral properties, and spatial singularities.

This article is organized as follows. Section 2 studies the wave propagation in a waveguide partially filled with ferrite slab adjacent to an additional air layer. We show that, as the thickness of this air layer is reduced to zero, unidirectional propagation is enforced. Section 3 develops a transmission-line model for the closed cavity. Our analysis shows that, under the same limiting process, the cavity spectrum evolves from discrete to continuous. Furthermore, the electromagnetic fields become increasingly localized, with the magnetic field strongly enhanced. Section 4 explores an alternative limiting procedure based on a nonlocal response model for the ferrite material. Section 5 discusses why the fields generated by an external source inside the cavity can be regular, while at the same time the normal modes do not have finite energy. Finally, section 6 summarizes the main findings.

## 2. Dispersion of a Ferrite-Loaded Waveguide

We consider a rectangular metallic waveguide partially filled with a ferrite slab of thickness $l$. The ferrite slab is placed between two air layers of thickness $\Delta$ on the left and $d$ on the right, as illustrated in Fig. 1a. The waveguide cross-section is $a \times b$, with $a$ the largest transverse dimension. The ferrite material is biased by a static magnetic field $\mathbf{H}_0$ aligned with the +$z$-direction, resulting in a gyrotropic permeability tensor of the form [1]:

$$\bar{\mu} = \mu_0 \begin{pmatrix} \mu_t & -i\mu_g & 0 \\ i\mu_g & \mu_t & 0 \\ 0 & 0 & 1 \end{pmatrix}, \qquad \mu_t = 1 + \frac{A\omega_0\omega_m}{\omega_0^2 A^2 - \omega^2}, \qquad \mu_g = \frac{\omega\omega_m}{\omega_0^2 A^2 - \omega^2}. \qquad (1)$$

Here, $\omega_0$ is the Larmor frequency determined by the applied magnetic bias $H_0$, $\omega_m$ is the Larmor frequency for the saturation magnetization, $A = 1 - i\alpha\, \omega / |\omega_0|$, and $\alpha > 0$ controls the strength of the damping due to material absorption. The electric response of the ferrite is described by a scalar permittivity $\varepsilon_0 \varepsilon_r$. The time variation $e^{-i\omega t}$ is implicit.

We focus on the transverse-electric modes TE$_{m0}$, , where $m$ is the mode index, for which $\mathbf{E} = E_z \hat{\mathbf{z}}$ and $\mathbf{H} = H_x \hat{\mathbf{x}} + H_y \hat{\mathbf{y}}$ and $\partial/\partial z = 0$. Their dispersion relation is [1, Sect. 9.3]:

$$\left(\frac{k_{\mathrm{f}}}{\mu_{\mathrm{ef}}}\right)^2+\left(\frac{\mu_g\beta}{\mu_t\mu_{\mathrm{ef}}}\right)^2-k_{\mathrm{a}}\cot\left(k_{\mathrm{a}}\Delta\right)\left[\frac{k_{\mathrm{f}}}{\mu_{\mathrm{ef}}}\cot\left(k_{\mathrm{f}}t\right)+\frac{\mu_g\beta}{\mu_t\mu_{\mathrm{ef}}}\right]$$
$$-\left(k_{\mathrm{a}}\right)^2\cot\left(k_{\mathrm{a}}\Delta\right)\cot\left(k_{\mathrm{a}}d\right)-k_{\mathrm{a}}\cot\left(k_{\mathrm{a}}d\right)\left[\frac{k_{\mathrm{f}}}{\mu_{\mathrm{ef}}}\cot\left(k_{\mathrm{f}}t\right)-\frac{\mu_g\beta}{\mu_t\mu_{\mathrm{ef}}}\right]=0, \quad (2)$$

where $\mu_{\mathrm{ef}}=\left(\mu_t^2-\mu_g^2\right)/\mu_t$, $k_{\mathrm{f}}=\sqrt{\mu_{\mathrm{ef}}\varepsilon_r\left(\omega/c\right)^2-\beta^2}$, $k_{\mathrm{a}}=\sqrt{\omega^2\mu_0\varepsilon_0-\beta^2}$ and $\beta$ is the guided propagation constant along the *x*-direction.

Figure 1b shows the dispersion relation $\omega(\beta)$ obtained using Eq. (2) for different values of the left air-region thickness, $\Delta$. The black, green and blue colors in the plot correspond to $\Delta=0$, $\Delta=0.001a$ and $\Delta=0.01a$, respectively. In this plot, we consider a lossless ferrite material corresponding to a damping factor $\alpha=0$. As seen in Fig. 1bi), the ferrite waveguide is strictly unidirectional when the left air layer width vanishes [9, 11, 23]. This behavior is generally observed in the spectral range $\left|\omega_0+\frac{\omega_m}{2}\right|\leq\omega\leq\left|\omega_0+\omega_m\right|$ [23].

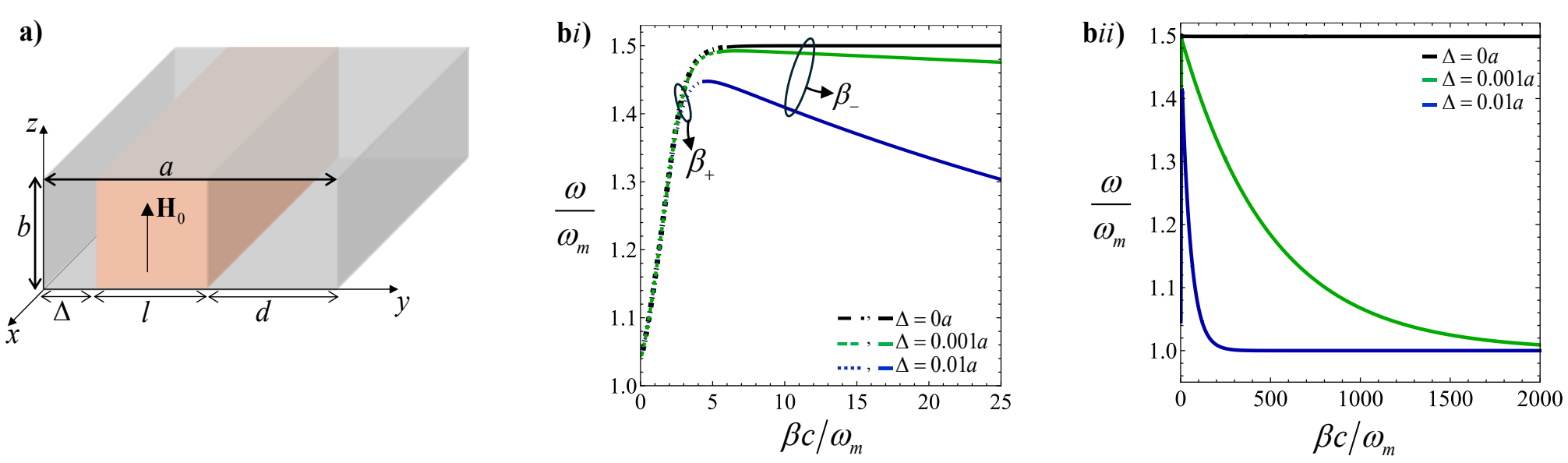

**Fig. 1 (a)** Geometry of a rectangular waveguide with cross-section $a\times b$ partially filled with a magnetized ferrite. The ferrite slab has thickness $l$ and is placed between air layers with thicknesses $\Delta$ and $d$. **(b*i*)** Dispersion relation $\omega(\beta)$ of the fundamental mode of a rectangular waveguide for different values of the left air layer thickness: $\Delta=0$, $\Delta=0.001a$ and $\Delta=0.01a$. The ferrite parameters are $\varepsilon_r=10$, $\omega_m=c/a$ and $\omega_0=0.5\omega_m$ and $\alpha=0$, and the right air layer thickness is $d=0.5a$. The part of the dispersion associated with a positive group velocity along +*x* is represented with discontinuous lines, whereas the part of the dispersion associated with a negative or null group velocity is represented with continuous lines. For $\Delta=0$ the guide is strictly unidirectional in the relevant spectral range. **(b*ii*)** Same as in bi) but the horizontal axis scale is expanded to $0<\beta c/\omega_m<2000$.

Interestingly, for any finite air-gap thickness $\Delta$ the waveguide becomes bidirectional and supports a forward-propagating mode (along +*x*, labeled $\beta_+$) and a backward propagating mode (along −*x*, labeled $\beta_-$). As $\Delta$ decreases toward zero, the $\beta_-$ branch flattens out, implying that the backward-mode group velocity approaches zero. Furthermore, for a finite $\Delta$ the backward mode spans the entire spectral range over which the forward mode propagates. This behavior is illustrated in Fig. 1bii), which shows that the crossing occurs in a spectral region on the order of $0\leq\beta\leq 2/\Delta$. This indicates that the counter-directional mode can acquire a very large propagation constant, consistent with strong spatial localization. The ability of an air gap to suppress localization has been noted in previous works [9, 17, 19]. Furthermore, it has been shown that the air gap mimics a wave-vector cutoff in the ferrite response and can restore the bulk-edge correspondence in the continuum [24, 25].

## 3. Nonreciprocal Cavity and Transmission-line Equivalent Model

Next, we consider the rectangular waveguide of Fig. 1a terminated by metallic walls placed at $x = 0$ and $x = L$, forming a cavity. This system can be modeled as a nonreciprocal transmission line short-circuited (s.c.) at both ends, as illustrated in Fig. 2a.

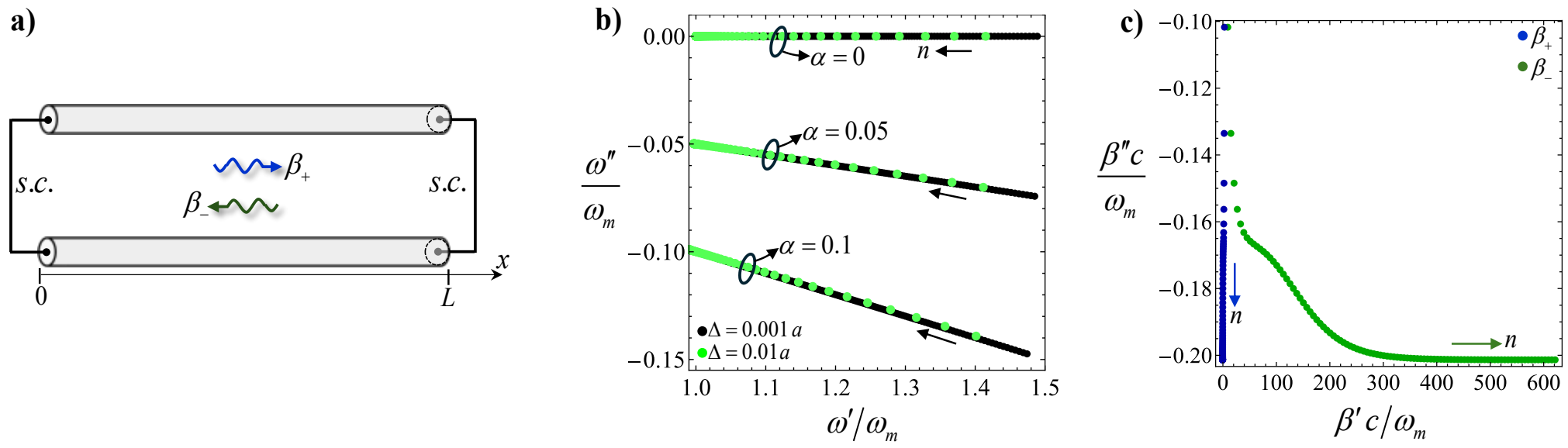

**Fig. 2 (a)** Transmission-line model of the ferrite loaded metallic cavity. The transmission line has length $L$ and is terminated in a short circuit (s.c.). The forward-propagating wave ($x$>0) is characterized by the propagation constant $\beta_+$ whereas the backward-propagating wave is defined by the propagation constant $\beta_-$. **(b)** Cavity frequencies $\omega = \omega' + i\omega''$ represented in the complex plane for different values of the left air-layer thickness: $\Delta = 0.01a$ (green dots) and $\Delta = 0.001a$ (black dots), and different values of the damping factor: $\alpha = 0$, $\alpha = 0.05$ and $\alpha = 0.1$. The waveguide parameters are the same as in Fig. 1b and the waveguide length is $L = a$. **(c)** Representation of $\beta_+ = \beta'_+ + i\beta''_+$ and $\beta_- = \beta'_- + i\beta''_-$ in the complex plane, with the propagation constants evaluated for the eigenmodes. The left air-layer has thickness of $\Delta = 0.01a$ and the ferrite damping factor is $\alpha = 0.1$. The arrows in the panels b) and c), indicate the direction of increasing cavity-mode order ***n*****.**

The electromagnetic field inside the cavity is approximated by the superposition of two counter-propagating traveling waves. The forward-propagating wave is characterized by the propagation constant along the +*x* direction, $\beta_+$, and the backward-propagating wave by the propagation constant along the –*x* direction, $\beta_-$. The total electric field distribution along the cavity is $E(x) = E_0^+ e^{i\beta_+ x} + E_0^- e^{i\beta_- x}$, where $E_0^+, E_0^-$ represent the amplitudes of the forward and backward waves, respectively. For simplicity, here we neglect the transverse variation of the field, which for nonreciprocal systems depends on the direction of propagation of the wave.

Due to the perfectly conducting walls at *x*=0 and *x*=*L*, the electric field is required to vanish at these boundaries. Enforcing $E|_{x=0} = 0$, one finds $E(x) = E_0^+ \left(e^{i\beta_+ x} - e^{i\beta_- x}\right)$. Enforcing the second boundary condition, $E|_{x=L} = 0$, one obtains the dispersion equation for the relevant cavity modes (specifically for the modes that arise from the interference of TE$_{m0}$ waveguide modes) within the transmission line approximation:

$$\beta_-(\omega) - \beta_+(\omega) = \frac{2\pi}{L} n, \tag{3}$$

where $n$ labels the mode number in the cavity, $n$ = 0,1,2,3…, and $\beta_+, \beta_-$ are obtained from the solutions of Eq. (2), as explained in the previous section. This simple transmission-line model makes explicit that the cavity spectrum is governed by the accumulated phase of the forward and backward waves. Therefore, when the backward branch flattens in the unidirectional limit, the usual discrete cavity spectrum is expected to collapse into a continuum.

This result is confirmed by Fig. 2b, which shows the normal-mode frequencies $\omega' + i\omega''$ of the cavity in the complex plane for different levels of ferrite dissipation and different air-gap thicknesses. Note that the real part $\omega'$ of the eigenfrequency determines the oscillation period, whereas the imaginary part $\omega''$, with $\omega'' < 0$, describes the lifetime of the oscillation due to dissipation. The eigenfrequencies are obtained numerically as the solutions of a coupled system of three equations: the modal equation (3) and the two dispersion relations for the forward and backward branches, $\beta_+(\omega)$ and $\beta_-(\omega)$, defined implicitly by Eq. (2). In Fig. 2b, the green dots represent the resonant frequencies of a cavity with $\Delta = 0.01a$, whereas the black dots are calculated for a cavity with a thinner air layer of thickness $\Delta = 0.001a$. The arrows in the figure indicate the direction of increasing cavity-mode order *n*. As shown, the density of modes increases as the thickness $\Delta$ tends to zero, and, in the $\Delta = 0^+$ limit, the eigenfrequencies $\omega'$ densely fill the entire spectral range, signaling the transition from a discrete to a continuous spectrum.

The origin of this effect is associated with the bending and flattening of the backward-wave dispersion curve [Fig. 1b] as the thickness decreases, which suppresses backscattering and drives the system into a unidirectional regime. Hence, the transition from bidirectional to unidirectional propagation is accompanied by a transition of the modal spectrum from discrete to continuous. In both cases ($\Delta = 0.01a$ and $\Delta = 0.001a$), in Fig. 2b, the accumulation of resonant frequencies is particularly pronounced near $\omega'/\omega_m \to 1$, because $\beta_-(\omega)$ approaches asymptotically infinity when $\omega \to \omega_m$ [Fig. 1b].

The main impact of material losses on the previous discussion is that dissipation shifts the eigenfrequencies from the real-frequency axis (set of points associated with $\alpha = 0$ in Fig. 2b, corresponding to a lossless system) toward the lower-half frequency plane (see the points associated with $\alpha = 0.05$ and $\alpha = 0.1$ in Fig. 2b; note that realistic levels of loss in ferrites can be as low as $\alpha = 10^{-4}$). However, independently of the dissipation level, the spectrum becomes continuous in the $\Delta = 0^+$ limit. This property can be attributed to the fact that ordinary cavity modes are formed by the interference of counter-propagating waves. In the unidirectional limit, this mechanism is no longer available, and adding dissipation to the system does not restore it.

Figure 2c shows the effect of material loss on the propagation constants of the forward and backward propagating waves, $\beta_+ = \beta'_+ + i\beta''_+$ and $\beta_- = \beta'_- + \beta''_-$, respectively, for $\alpha = 0.1$. The propagation constants are evaluated at the cavity eigenfrequencies. For a fixed mode, the imaginary parts of the forward and backward waves, $\beta_+, \beta_-$, are identical, i.e. $\beta''_+ = \beta''_- \equiv \beta''$. This property follows from Eq. (3). As seen in Fig. 2c, $\beta''$ remains uniformly bounded in the entire spectral region. Furthermore, the result of detailed simulations indicates that $|\beta''|_{\max} \sim \alpha\,\omega_m / c$, i.e., the strength of the imaginary part of the propagation constants is uniformly bounded by the ferrite dissipation level, and is largely independent of the air-gap thickness $\Delta$. In contrast, the real part of the propagation constant of the backward mode ($\beta'_-$) scales approximately as $1/\Delta$ near a fixed frequency $\omega'$ in most of the spectral window where the waveguide is unidirectional.

Next, we investigate how the gap width influences the fields in the cavity. To this end, we further develop the transmission-line analogy used in the preceding analysis. We begin by discussing the equivalent voltage $V$. As the electric field is oriented along *z*, it may be defined as $V = -\int_0^b \langle E_z \rangle dz = -b\langle E_z \rangle$, where $\langle E_z \rangle = \frac{1}{a}\int_0^a E_z dy$ denotes spatial averaging along the *y*-direction and *b* is the cross-section dimension along *z*. We used the fact that the fields are independent of *z* for the considered family of waveguide modes. Likewise, the equivalent current $I$ is defined as the current on the waveguide plate $z = b$. The surface current density on this waveguide plate is $\mathbf{K}=\hat{\mathbf{n}}\times\mathbf{H}$ with $\hat{\mathbf{n}}=-\hat{\mathbf{z}}$ is the unit vector normal to the upper cavity wall. Thus, the current flowing along $x$ is determined by $I=\int_0^a H_y\,dy$.

Consider now that the voltage and current are associated with one of the waveguide modes, i.e., a traveling wave with propagation factor $e^{i\beta x}$. The previous considerations then show that the associated characteristic impedance of the equivalent line is:

$$Z_c=-\frac{b}{a}\frac{\langle E_z\rangle}{\langle H_y\rangle}=\omega\mu_0\frac{b}{a}\frac{\langle E_z\rangle}{\left\langle\frac{1}{\mu_{\rm ef}}\left(\beta E_z+\frac{\mu_{\rm g}}{\mu_{\rm t}}\frac{\partial E_z}{\partial y}\right)\right\rangle}, \tag{4}$$

where $\langle...\rangle$ denotes spatial averaging along the $y$-direction. To obtain the rightmost identity, we used Faraday's law: $\nabla\times\mathbf{E}=i\omega\overline{\mu}\cdot\mathbf{H}$. It is understood that the permeability elements vary in space such that in the air regions we have $\mu_g=0$ and $\mu_{\rm ef}=\mu_t=1$.

To proceed and obtain an analytical expression, we make a few approximations. First, we neglect the term involving $\mu_g$ in the expression for the characteristic impedance. Second, we use the well-known property of ferrite waveguides that the forward mode has most of its energy in the right air region of the guide, where $\mu_{\rm ef}=1$, whereas the backward mode is mainly concentrated in the ferrite region, where $\mu_{\rm ef}$ is evaluated using the ferrite permeability [1, 9, 11]. These approximations lead to the following explicit expressions for the characteristic impedances of the forward and backward waves:

$$Z_c^+\approx\frac{\omega\mu_0}{\beta_+}\frac{b}{a}, \qquad Z_c^-\approx\frac{\omega\mu_0\mu_{\rm ef}}{\beta_-}\frac{b}{a}. \tag{5}$$

As seen, the two characteristic impedances are different. For waves with $\omega$ on the order of $\omega_m$, their ratio can be roughly estimated as $\frac{Z_c^-}{Z_c^+}=\mu_{\rm ef}\frac{\beta_+}{\beta_-}\sim\mu_{\rm ef}\frac{\omega_m\Delta}{c}$, i.e., the impedance ratio approaches zero when the air-gap is closed.

Within this model, the voltage and current in the cavity take the form

$$V(x)=V_0^+e^{i\beta_+x}+V_0^-e^{i\beta_-x}, \qquad I(x)=\frac{V_0^+}{Z_c^+}e^{i\beta_+x}+\frac{V_0^-}{Z_c^-}e^{i\beta_-x}. \tag{6}$$

Note that in our system both $\beta_+,\beta_-$ are positive, but as in the spectral range of interest the ferrite effective permeability is negative ($\mu_{\rm ef}<0$), the impedance $Z_c^-$ is also negative. This ensures that the power carried by backward wave flows along $-x$, as expected. Enforcing the boundary condition $V|_{x=0}=0$, the voltage and current in the equivalent line reduce to:

$$V(x)=V_0^+\left(e^{i\beta_+x}-e^{i\beta_-x}\right), \qquad I(x)=\frac{V_0^+}{Z_c^+}\left(e^{i\beta_+x}-\frac{\beta_-}{\mu_{\rm ef}\beta_+}e^{i\beta_-x}\right). \tag{7}$$

Figure 3 depicts the spatial profile of the magnitude of the equivalent voltage and current calculated for the mode with $\omega'\approx1.128\omega_m$. This mode is representative: the results for other frequencies are qualitatively similar. The voltage and current distributions in the resonator are weakly sensitive to $\alpha$. For nonzero $\alpha$, the fields tend to concentrate at one end of the cavity because the counterpropagating modes experience different levels of attenuation.

Comparing the upper and lower panels of Fig. 3a, one sees that for a fixed $\omega'$ the number of voltage oscillations and nulls increases when $\Delta$ approaches zero. In contrast, Fig. 3b shows that the oscillatory behavior observed in the voltage is strongly suppressed in the current for small $\Delta$. In addition, the magnitude of the current is strongly sensitive to variations in $\Delta$ scaling roughly as $1/\Delta$. This behavior results from the strong asymmetry between the characteristic impedances of the counter-propagating transmission-line modes. It implies that the magnetic-field distribution becomes singular during the transition from the bidirectional to the unidirectional regime, even in the presence of strong dissipation.

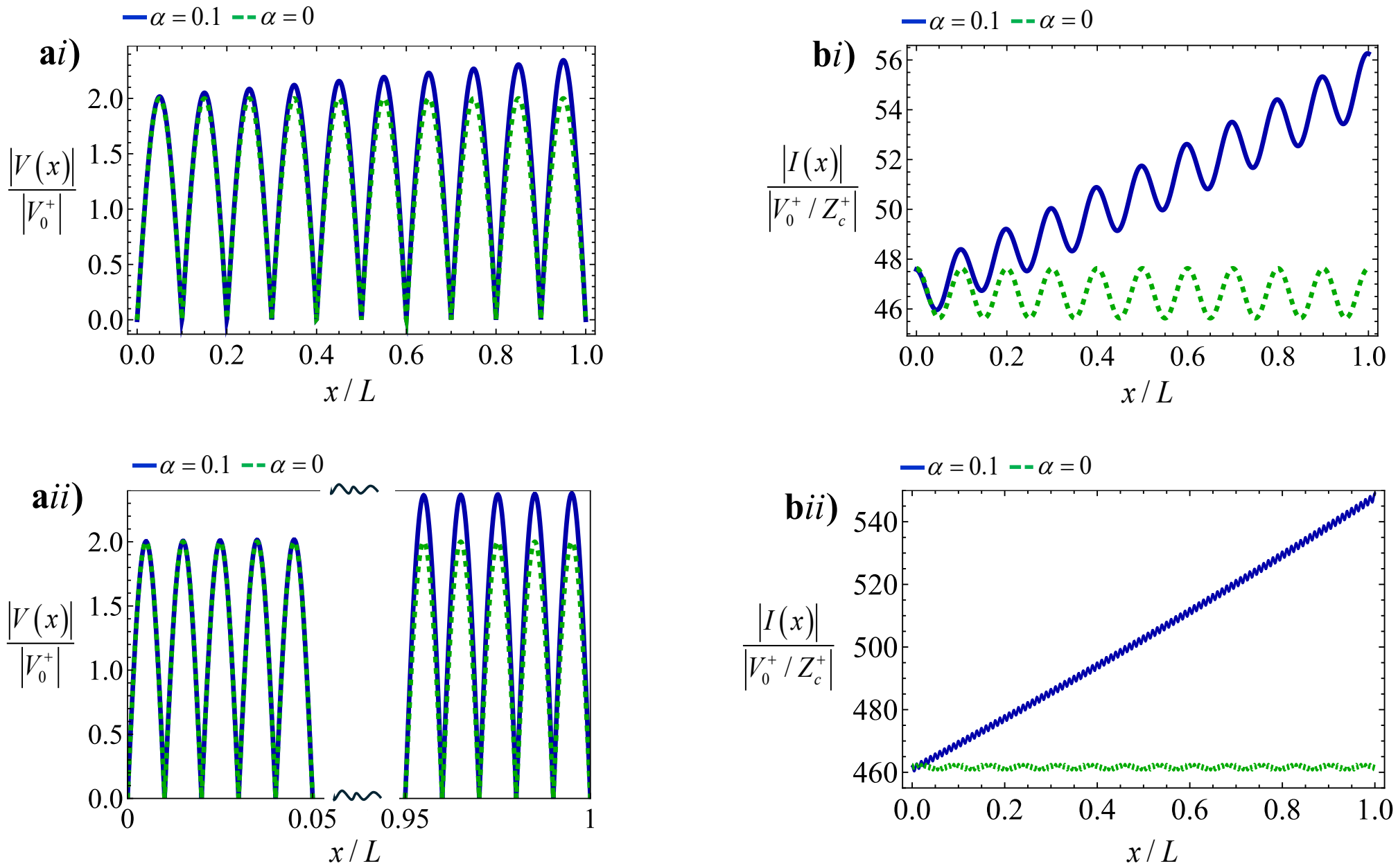

**Fig. 3 (a)** Spatial profile of the normalized voltage magnitude $|V(x)|$ along the *x*-direction of the transmission-line cavity, for the resonant mode with $\omega' \approx 1.128\omega_m$. The ferrite parameters are the same as in Fig. 1b, with $d = 0.5a$. The transmission line length is $L = a$. The solid green curves are computed for a damping factor of $\alpha = 0$ (lossless case) whereas the solid blue curves correspond to a damping factor of $\alpha = 0.1$. The upper row panel is obtained for a left air-layer thickness of $\Delta = 0.01a$ and the lower row panel for $\Delta = 0.001a$. Note that in the lower panel the central region of the plot is not shown. **(b)** Similar to (a) but for the normalized current magnitude $|I(x)|$.

This property can be understood qualitatively as follows. In conventional cavities, the normal modes are standing waves formed by the interference of two counter-propagating waves, so that each wavefront repeatedly travels back and forth between the two ends of the cavity before it is eventually dissipated. In a strictly unidirectional system, however, the return channel is absent, and any free-field solution must therefore exhibit singular behavior [23]. Despite its simplicity, the transmission-line model captures this singularity by predicting that the current, and hence the magnetic field in the cavity, diverges in the $\Delta = 0^+$ limit. The divergence of the magnetic field at specific points in the cavity was demonstrated in Ref. [23] using a quasi-static approach. Specifically, it was shown that, in the presence of dissipation, any normal mode that propagates along the positive *x*-direction must retain singular behavior near $x$=0 [23]. This source-type singularity is an inevitable consequence of the absence of backscattering.

To further validate our transmission-line model, we simulated the effect of changing the air-gap width in the nonreciprocal ferrite cavity using CST Studio Suite [26]. Specifically, we excited the cavity with a broad Gaussian pulse and recorded the electric field along the longitudinal direction of the waveguide ($x$-direction) as a function of time. Then, by applying Fourier transforms in time and along the longitudinal direction, we resolved the cavity response in both frequency and wave vector ( $\beta$ ). The simulation results are shown in Fig. 4 for $\Delta = 0.2a$ and $\Delta = 0.01a$. The bright spots in the plots indicate the cavity modes. Typically, for a given frequency, there are two bright spots, corresponding to the forward- and backward-propagating modes of the transmission-line model. Figure 4b shows that, for small $\Delta$, the spectrum is nearly continuous and almost coincides with the dispersion of the associated waveguide mode. In contrast, for $\Delta = 0.2a$, the cavity supports only a few modes. This illustrates how displacing the ferrite block along the waveguide cross-section drives the transition of the cavity spectrum from a discrete to a quasi-continuous regime.

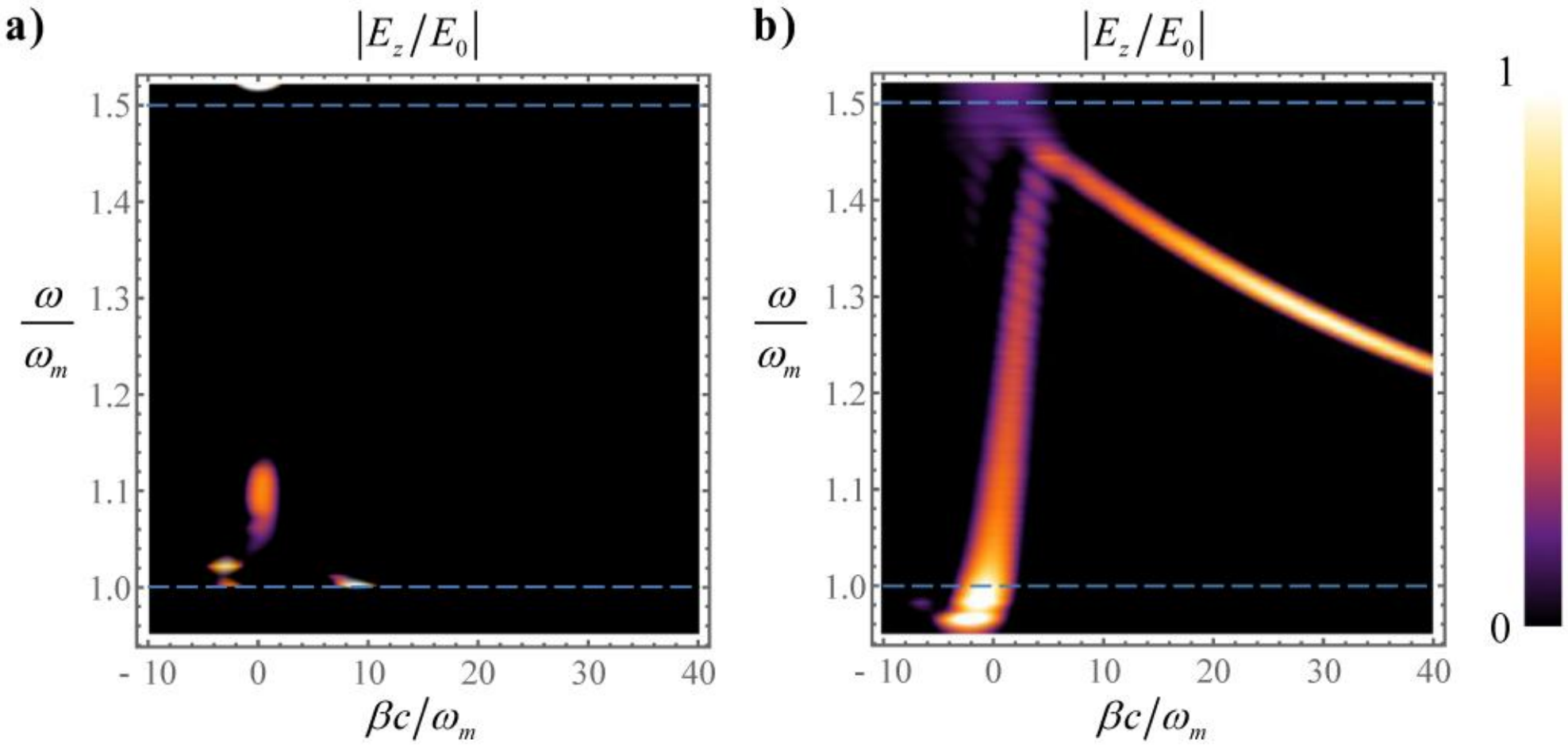

**Fig. 4** Normalized cavity spectrum $|E_z(\omega,\beta)/E_0|$ obtained with CST Studio Suite [26] for **(a)** $\Delta = 0.2a$, and **(b)** $\Delta = 0.01a$. The structural parameters are as in Fig. 1b, except that the ferrite thickness is $l = 0.5a$, the guide length is $L = 2a$ and dissipation level is $\alpha = 0.002$. The waveguide cross-section is such that $b/a$=0.66. The horizontal dashed lines delimit the spectral window for which the guide becomes unidirectional in the limit $\Delta = 0^+$.

## 4. Ferrite with a Non-local response

Next, we discuss the impact of nonlocality in the ferrite response on the dispersion characteristics of the guided mode. In ferrimagnetic materials, this nonlocality is usually attributed to the spin-exchange interaction, which tends to keep neighboring spins aligned over short length scales set by the exchange length $l_{\text{ex}}$ [27, 28]. This effect can be modeled by considering that the effective magnetic field associated with the spin torque in the Landau-Lifshitz-Gilbert equation has an additional component, the exchange field, such that $\mathbf{H}_{\text{ef}} = \mathbf{H} + l_{\text{ex}}^2 \nabla^2 \mathbf{M}$ [27, 28]. We show in Appendix A that this leads to a modified permeability response, which retains the gyrotropic structure of Eq. (1), but in which the Larmor frequency becomes wave-vector dependent: $\omega_0 \to \omega_{0,k} = \omega_0 + \omega_m l_{\text{ex}}^2 k^2$ [27, 28]. We note in passing that the role of the exchange interaction in the magnetic response is qualitatively similar to that of the odd-viscosity term considered by Vitelli and co-workers in gyrotropic continuum models [29]. For simplicity, in this section we neglect the effect of dissipation, as the lossless case is typically the most demanding from the standpoint of response regularization.

In Appendix B, we develop the electromagnetic theory for the waveguide modes of the partially filled ferrite guide of Fig. 1, including nonlocal effects. This involves writing the waveguide fields in terms of bulk plane waves in each homogenous region and enforcing suitable boundary conditions at the interfaces and metallic walls. Due to the nonlocality of the ferrite, it is necessary to enforce an additional non-Maxwellian boundary condition at the ferrite boundaries, which imposes that the normal derivative of the magnetization vector vanishes [28]. This procedure yields a homogeneous system of linear equations whose determinant determines the dispersion equation for the $TE_{m0}$ modes. We focus on the most challenging scenario, in which the left air layer is suppressed ( $\Delta = 0$ ).

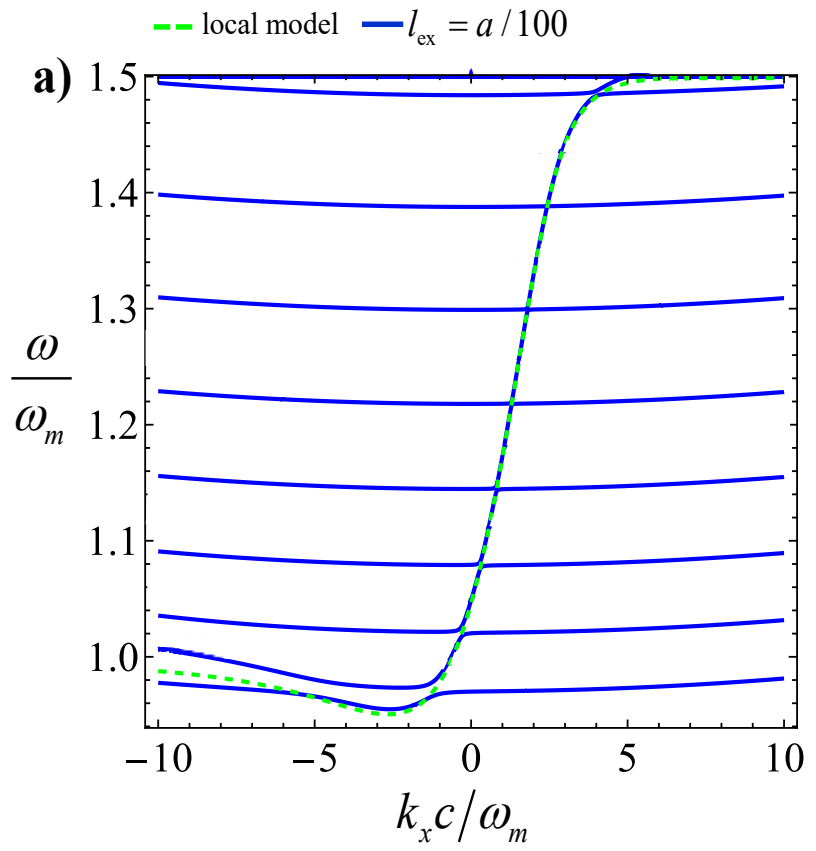

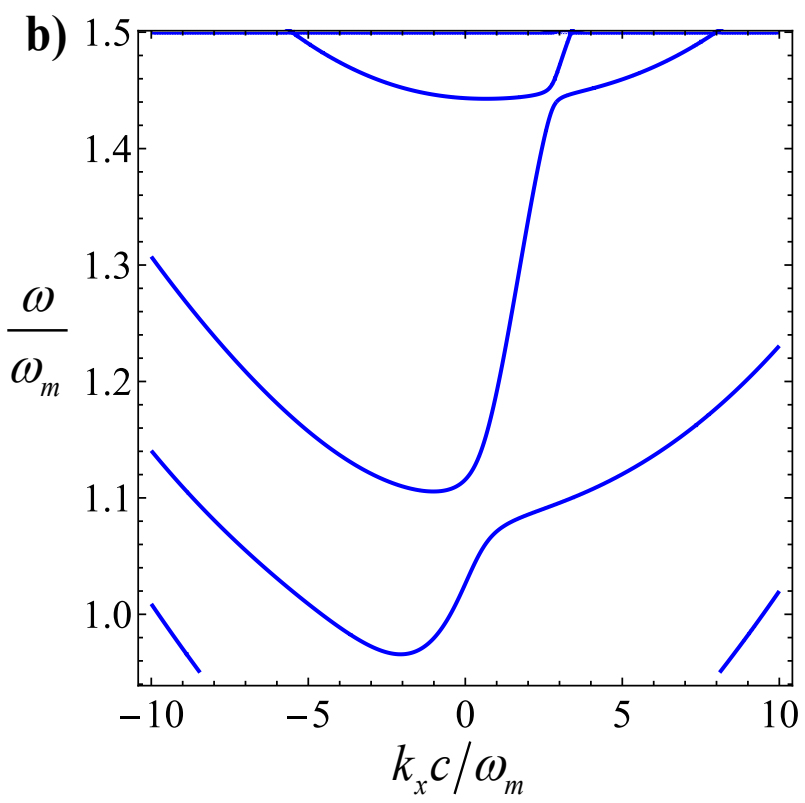

**Fig. 5** Dispersion of the fundamental mode of a waveguide partially filled with a nonlocal ferrite material for the configuration of Fig.1a with $\Delta = 0$ . The ferrite slab has thickness $l = 0.5a$ and the constitutive parameters are the same as in Fig. 1b, except for the exchange length $l_{ex}$ which sets the strength of the nonlocal response. **(a)** Blue line: $l_{ex} = a/100$ ; Green line: local model ( $l_{ex} = 0$ ). **(b)** $l_{ex} = a/25$ .

Figure 5 depicts the modal dispersion for a waveguide with the same structural parameters as in Fig. 1b and $\Delta = 0$ . The nonlocality in the ferrite is governed by the exchange length $l_{ex}$ . In typical ferrimagnetic materials, $l_{ex}$ is on the order of a few to several tens of nanometers [30]. Hence, for a microwave waveguide where $a \sim \text{cm}$ , a realistic estimate for the normalized exchange length is $l_{ex} \sim a/10^7$ , which is very small on the electromagnetic length scale. To illustrate more clearly the impact of nonlocality, the numerical examples use exchange lengths several orders of magnitude larger than realistic values. This not only simplifies the visualization but also facilitates the numerical extraction of the dispersion curves, since for very small $l_{ex}$ the dispersion diagram contains many branches that are difficult to resolve numerically.

As seen in Fig. 5 (blue lines), a finite $l_{ex}$ renders the system bidirectional, similar to the effect of having a finite-thickness air slab, $\Delta \neq 0$ , in the local-response model discussed in Sect. 2. Thus, the exchange interaction suppresses the problematic unidirectional behavior characteristic of the local response limit (green dashed line in panel a)). In the nonlocal model, smaller values of the exchange length [compare Fig. 5a with Fig. 5b] bring the system closer to the unidirectional regime. This is accompanied by the opening of many small gaps in the original unidirectional-mode dispersion, as well as by an increasingly large number of bi-directional dispersion branches with progressively smaller group velocity. In the limit, $l_{\mathrm{ex}} \to 0^+$ the nonlocal model yields the original unidirectional mode of the local dispersion, together with many flat bands for which the group velocity of the wave is exactly zero.

Since for any finite $l_{\mathrm{ex}}$ the system is bidirectional and has no flat bands, the spectrum of a cavity formed by the waveguide partially filled with the nonlocal ferrite is expected to be discrete. A closely related regularization mechanism has been reported in singular plasmonic systems [31-34], where nonlocal effects arising from diffusion suppress the continuous spectrum predicted by the local-response model and replace it with a discrete set of resonances [34].

## 5. Discussion

As shown in the previous sections, a dissipative ferrite cavity develops a continuous spectrum in the limit where the waveguide becomes strictly unidirectional. Furthermore, it was shown in Ref. [23] that in such regime the waveguide modes necessarily exhibit singular, non-square-integrable behavior, even in the presence of dissipation, due to the absence of a conventional standing-wave pattern. Indeed, a closed strictly one-way system cannot sustain a regular source-free modal field without a singularity.

At the same time, numerical simulations reported in several previous works show that material dissipation can be sufficient to ensure a well-defined electromagnetic response in driven problems [12-17], when the cavity is excited by an external source, without any apparent singular behavior. This raises the question as to why the driven problem is regular, whereas the spectral problem exhibits problematic modes, especially because the Green function may be expressed in terms of the system modes. Indeed, the Green's function of a non-Hermitian system can generally be expanded in terms of a biorthogonal basis of electromagnetic modes as follows [23, 35]:

$$\overline{\mathcal{G}}\left(\mathbf{r},\mathbf{r}',\omega\right)=\frac{\omega}{2}\sum_{n}\frac{1}{\omega_n-\omega}\mathbf{f}_n\left(\mathbf{r};\alpha\right)\otimes\mathbf{f}_n^*\left(\mathbf{r}';-\alpha\right). \tag{8}$$

Here $\mathbf{f}_n\left(\mathbf{r};\alpha\right)=\left(\mathbf{E}_n \quad \mathbf{H}_n\right)^T$ represent the modes of the dissipative cavity, whereas $\mathbf{f}_n\left(\mathbf{r};-\alpha\right)$ represent the modes of the adjoint problem, in which material dissipation is replaced by gain [23]. The above expression suggests that any singularities in $\mathbf{f}_n\left(\mathbf{r};\alpha\right)$ should be inherited by the Green's function, and hence by the fields generated in a driven problem.

The resolution of this apparent contradiction lies in the fact that the cavity spectrum becomes continuous in the strictly unidirectional regime. In that case, the modal summation must be replaced by a spectral integral over generalized eigenstates. Although the individual modes may exhibit singular behavior and fail to be square integrable, the singularities can be regularized upon integration over the continuous spectrum. Thus, the Green function can remain regular even though the individual eigenstates entering the spectral representation are not square integrable. This is a standard feature of systems with continuous spectra. A familiar example is free space, whose spectral representation is built from plane waves. The individual plane waves do not have finite energy, but superpositions of them, such as wave packets or fields generated by localized sources, can be square-integrable.

The continuous spectrum is also relevant in the discussion of the failure of the conventional fluctuation-dissipation-theorem (FDT) description based on the local material model, as well as the thermodynamic paradox discussed in Ref. [23]. Indeed, FDT correlations involve the anti-Hermitian part of the Green function evaluated at coincident observation and source points, $\mathbf{r}=\mathbf{r}'$, which is a much more singular operation than computing the driven response to a distributed external source [23, 36]. Therefore, the singular behavior of the eigenstates may not be suppressed by the integration over the continuous spectrum when one sets $\mathbf{r}=\mathbf{r}'$. Thus, even if the driven problem is regular, the FDT local correlations and the corresponding equilibrium Poynting vector $\mathbf{S}_\omega$ may remain singular, leading to the apparent thermodynamic paradox discussed in Ref. [23]. It should be noted that the paradox does not imply a failure of the FDT formalism itself, but rather that the idealized local material model inserted into the FDT calculation is physically incomplete. Material loss alone does not necessarily provide the missing physical cutoff.

## 6. Conclusion

We have shown that a cavity formed by closing a strictly unidirectional waveguide is characterized by a continuous spectrum. In the absence of backscattering, the usual standing-wave mechanism is lost, and the discrete set of cavity resonances collapses into a continuum. Our analysis is based on a simple transmission-line model, which also captures the accompanying field singularity through the divergence of the equivalent current, and hence of the magnetic field. The discrete-to-continuous spectral transition was demonstrated in a ferrite-loaded microwave cavity using two complementary regularization schemes. In the first, we show that a finite air gap restores bidirectional propagation and a discrete spectrum. In the second, exchange-induced nonlocality in the ferrite response provides a material cutoff that also restores a discrete spectrum for any finite exchange length.

Our results clarify why lossy driven simulations can appear well behaved even though the corresponding source-free modes are singular. Dissipation can make the Green's operator regular when acting on distributed sources, but it does not by itself restore cavity modes with finite energy or a discrete spectrum. Furthermore, the singularities of the cavity modes may still influence the correlations of fields generated by thermal agitation. These correlations probe the Green's function evaluated at $\mathbf{r}=\mathbf{r}'$ and may therefore remain sensitive to the singular continuous-spectrum limit. Thus, material loss alone is not a sufficient regularization mechanism for ideal nonreciprocal cavities: a physical cutoff, such as finite geometry or nonlocal material response, is required.

**Epilogue:** We dedicate this article to Professor Nader Engheta on the occasion of his 70th birthday. Professor Engheta has long been a mentor and a source of inspiration, and has shaped our scientific paths through his enthusiasm, generosity, and distinctive taste for bold ideas. This article is, in many ways, a testament to his influence. One of us, M.S., was first exposed to the problem of nonreciprocal wave propagation in waveguides, and to the possibility of using one-way propagation for field concentration, during a sabbatical stay at the University of Pennsylvania in 2010. That experience opened the door to many subsequent scientific studies, including some of the questions addressed in this work. Happy birthday, Professor Engheta.

## Acknowledgments

This work is supported in part by the Institution of Engineering and Technology (IET), by the Simons Foundation (award SFI-MPS-EWP-00008530-10), and by national funds through FCT – Fundação para a Ciência e a Tecnologia, I.P., and, when eligible, co-funded by EU funds under project/support UID/50008/2025 – Instituto de Telecomunicações, with DOI identifier https://doi.org/10.54499/UID/50008/2025. D. E. F. acknowledges financial support by IT-Lisbon and FCT under a research contract Ref. CEECINST/00058/2021/CP2816/CT0003 and Grant DOI https://doi.org/10.54499/CEECINST/00058/2021/CP2816/CT0003.

## Appendix A: Electrodynamics of a Ferrite with a Non-Local Response

In this Appendix, we derive the linearized response of a ferrite including nonlocal corrections [27, 28]. The dynamics of magnetic dipole moments under an external magnetic field is governed by the Landau–Lifshitz-Gilbert equation:

$$\frac{d\mathbf{M}}{dt} = -\frac{e}{m}\mathbf{M}\times\mu_0\mathbf{H}_{\text{ef}} + \frac{\alpha}{M_s}\mathbf{M}\times\partial_t\mathbf{M} . \tag{A1}$$

For simplicity, we take the $g$-factor to be exactly 2. Here, $-e$ is the electron charge and $m$ the electron effective mass, $M_s$ is the saturation magnetization, $\mathbf{M}$ is the magnetization vector and $\mathbf{H}_{\text{ef}}$ is the effective magnetic field responsible for the spin torque. The second term on the right-hand side is the Gilbert damping term, which accounts for dissipative effects.

In standard descriptions, the effective magnetic field is identified with the external magnetic field $\mathbf{H}$. However, micromagnetic theory introduces a nonlocal correction associated with the spatial variation of the magnetization itself: $\mathbf{H}_{\text{ef}} = \mathbf{H} + l_{\text{ex}}^2\nabla^2\mathbf{M}$ [27, 28]. This correction is known as the exchange contribution and penalizes rapid spatial variations of the magnetization vector. The relevant length scale at which this correction becomes important is the exchange length, $l_{\text{ex}}$. The exchange length sets the scale over which magnetization textures can vary appreciably.

The dynamic response of a magnetically biased ferrite material in the presence of the exchange term can be found in the usual manner. Specifically, we write the magnetic field and magnetization as the sum of static and time-harmonic components, $\mathbf{H}\to\mathbf{H}_0 + \mathbf{H}_\omega e^{-i\omega t}$ and $\mathbf{M}\to\mathbf{M}_0 + \mathbf{M}_\omega e^{-i\omega t}$, and then linearize the Landau–Lifshitz-Gilbert equation with respect to the dynamic part of the signal. The result is:

$$-i\omega\mathbf{M}_\omega = -\boldsymbol{\omega}_m\times\left[\mathbf{H}_\omega + l_{\text{ex}}^2\nabla^2\mathbf{M}_\omega\right] - \mathbf{M}_\omega\times\boldsymbol{\omega}_0 - i\omega\alpha\frac{\boldsymbol{\omega}_m}{|\omega_m|}\times\mathbf{M}_\omega . \tag{A2}$$

We define $\boldsymbol{\omega}_0 = \frac{e}{m}\mu_0\mathbf{H}_0$ as the Larmor (cyclotron) frequency vector, and $\boldsymbol{\omega}_m = \frac{e}{m}\mu_0\mathbf{M}_{\text{s}}$ as the Larmor frequency for the saturation magnetization. Then, for a static bias along the $z$-direction:

$$-i\omega\mathbf{M} = -\omega_m\hat{\mathbf{z}}\times\mathbf{H} + \left(\omega_0 - i\omega\alpha\frac{\omega_m}{|\omega_m|}\right)\hat{\mathbf{z}}\times\mathbf{M} + \hat{\mathbf{z}}\times\left[-\omega_m l_{\text{ex}}^2\nabla^2\mathbf{M}\right]. \tag{A3}$$

We drop the subscripts $\omega$ in the magnetization and in the magnetic field, as from this point on these quantities refer to the dynamic fields. In the Fourier domain, for fields varying as $e^{i\mathbf{k}\cdot\mathbf{r}}$, we can take $\nabla^2 = -k^2$. This yields the familiar master equation for the ferrite magnetization, but with a wave-vector-dependent Larmor frequency:

$$-i\omega\mathbf{M} = -\omega_m\hat{\mathbf{z}}\times\mathbf{H} + \omega_{0,k}\hat{\mathbf{z}}\times\mathbf{M} , \qquad \omega_{0,k} = \omega_0 - i\omega\alpha\frac{\omega_m}{|\omega_m|} + \omega_m l_{\text{ex}}^2 k^2 . \tag{A4}$$

Hence, the linearized response of the ferrite with the "exchange" magnetic field term has the same structure as in Eq. (1) of the main text, but with nonlocal permeability elements given by:

$$\mu_t(\omega,k)=1+\frac{\omega_{0,k}\omega_m}{\omega_{0,k}^2-\omega^2}, \qquad \mu_g(\omega,k)=\frac{\omega\omega_m}{\omega_{0,k}^2-\omega^2}. \tag{A5}$$

For future reference, we note that the dispersion of the bulk transverse electric (TE) plane waves with $\mathbf{E}=E_z\hat{\mathbf{z}}$, $\mathbf{H}=H_x\hat{\mathbf{x}}+H_y\hat{\mathbf{y}}$ and $\partial/\partial z=0$, assumes the familiar form [1]:

$$\left(\frac{\omega}{c}\right)^2\mu_{\text{ef}}\left(\omega,k^2\right)\varepsilon_r=k^2, \tag{A6}$$

with $\mu_{\text{ef}}=\left(\mu_t^2-\mu_g^2\right)/\mu_t$ the effective permeability with the nonlocal correction.

## Appendix B: Dispersion Equation for a Waveguide Loaded with Non-Local Ferrite

In this Appendix, we derive the dispersion relation for the TE$_{m0}$ modes of the waveguide shown in Fig. 1 of the main text, incorporating the effect of the exchange interactions and the associated nonlocality in the ferrite response. For simplicity, we take the thickness of the left air layer to be identically zero, $\Delta=0$.

The structure of the modal fields is the same as in the main text: $\mathbf{E}=E_z\hat{\mathbf{z}}$ and $\mathbf{H}=H_x\hat{\mathbf{x}}+H_y\hat{\mathbf{y}}$ and $\partial/\partial z=0$. In particular, all field components can be expressed in terms of the electric field $E_z$. Since the waveguide cross section is inhomogeneous, with the ferrite occupying the region $0<y<l$, and the air layer occupying the complementary range $l<y<a$, we express $E_z$ as a linear combination of plane waves in each region and enforce the appropriate boundary conditions at the relevant boundaries. The field variation along $x$ is of the form $e^{ik_x x}$ with $k_x=\beta$ the propagation constant of the mode.

The air layer supports only two plane waves with the required structure: $e^{ik_x x}e^{\pm\gamma_0 y}$, where $\gamma_0=\sqrt{k_x^2-\omega^2/c^2}$. In contrast, due to the nonlocal effects, the bulk ferrite supports *six* plane waves with the required structure: $e^{ik_x x}e^{\gamma_{\text{f,n}}y}, e^{ik_x x}e^{-\gamma_{\text{f,n}}y}$, $n=1,2,3$, where $\gamma_{\text{f,n}}=\sqrt{k_x^2-k_n^2}$. Here, $k_n^2$ denotes one of the three solutions for $k^2$ of the bulk dispersion of the nonlocal ferrite [Eq. (A6)]: $\left(\frac{\omega}{c}\right)^2\mu_{\text{ef}}\left(\omega,k^2\right)\varepsilon_r=k^2$. The emergence of additional waves is a feature of both natural nonlocal materials and metamaterials [37, 38, 39].

The previous considerations lead to the following ansatz for the electric field of the waveguide modes:

$$E_z=E_0e^{ik_x x}\begin{cases}\sum_{n=1}^{3}A_{+,n}e^{-\gamma_{\text{f,n}}y}+A_{-,n}e^{+\gamma_{\text{f,n}}(y-l)}, & 0<y<l\\ A_0\dfrac{\sinh\left(\gamma_0(y-a)\right)}{\sinh\left(\gamma_0(l-a)\right)}, & l<y<a,\end{cases}. \tag{B1}$$

Here, $A_{+,n}$ and $A_{-,n}$ represent the amplitude of modes in the ferrite region and $A_0$ governs the amplitude of the modes in the air region. In writing the fields in the air region, we have already enforced the PEC boundary condition on the right wall, $E_z\big|_{y=a}=0$, which allows considering only a single wave amplitude in the air region. We choose the square root branch in $\gamma_{\text{f,n}}$ such that $\text{Re}\{\gamma_{\text{f,n}}\}\geq 0$.

The magnetic field distribution in the different regions can now be found using Maxwell's equations, taking into account that the electric field is a superposition of plane waves. Specifically, using Faraday's law, $\nabla\times\mathbf{E}=i\omega\bar{\mu}\cdot\mathbf{H}$, with $\nabla\times\mathbf{E}=\nabla E_z\times\hat{\mathbf{z}}$, and the structure of the nonlocal permeability tensor, which has the same form as in Eq. (1), one can show that the magnetic field associated with the $n$-th plane wave component is:

$$H_{x,n}=\frac{1}{i\omega\mu_0\left(\mu_{t,n}^2-\mu_{g,n}^2\right)}\left[\mu_{t,n}\partial_y E_{z,n}-i\mu_{g,n}\partial_x E_{z,n}\right], \tag{B2a}$$

$$H_{y,n}=\frac{1}{i\omega\mu_0\left(\mu_{t,n}^2-\mu_{g,n}^2\right)}\left[-i\mu_{g,n}\partial_y E_{z,n}-\mu_{t,n}\partial_x E_{z,n}\right]. \tag{B2b}$$

Here, $\mu_{t,n}$, $\mu_{g,n}$ are the elements of the nonlocal permeability tensor [Eq. (A5)] evaluated for the solution $k_n^2$ of the bulk ferrite dispersion [Eq. (A6)].

Furthermore, combining $\mathbf{M}=\left(\frac{\bar{\mu}}{\mu_0}-\mathbf{1}\right)\cdot\mathbf{H}$ with $\nabla E_z\times\hat{\mathbf{z}}=i\omega\bar{\mu}\cdot\mathbf{H}$, one can express the magnetization vector in terms of the electric field as: $\mathbf{M}=\frac{1}{i\omega\mu_0}\left(\mathbf{1}-\mu_0\bar{\mu}^{-1}\right)\cdot\left(\nabla E_z\times\hat{\mathbf{z}}\right)$. Thus, the magnetization vector associated with the $n$-th plane wave component is:

$$M_{x,n}=\frac{1}{i\omega\mu_0}\left[\partial_y E_{z,n}-\frac{1}{\left(\mu_{t,n}^2-\mu_{g,n}^2\right)}\left(\mu_{t,n}\partial_y E_{z,n}-i\mu_{g,n}\partial_x E_{z,n}\right)\right], \tag{B3a}$$

$$M_{y,n}=\frac{1}{i\omega\mu_0}\left[-\partial_x E_{z,n}-\frac{1}{\left(\mu_{t,n}^2-\mu_{g,n}^2\right)}\left(-i\mu_{g,n}\partial_y E_{z,n}-\mu_{t,n}\partial_x E_{z,n}\right)\right]. \tag{B3b}$$

Combining Eqs. (B2) and (B3) with Eq. (B1), we find the following explicit formulas for the magnetic field and magnetization distributions associated with a waveguide mode:

$$H_x=\frac{E_0}{i\omega\mu_0}e^{ik_x x}\begin{cases}\sum_{n=1}^{3}A_{+,n}\theta_n^{+}e^{-\gamma_{\mathrm{f,n}}y}+A_{-,n}\theta_n^{-}e^{+\gamma_{\mathrm{f,n}}(y-l)}, & 0<y<l\\ A_0\gamma_0\frac{\cosh\left(\gamma_0(y-a)\right)}{\sinh\left(\gamma_0(l-a)\right)}, & l<y<a,\end{cases}, \tag{B4}$$

and

$$M_x=\frac{E_0}{i\omega\mu_0}e^{ik_x x}\sum_{n=1}^{3}A_{+,n}\Theta_{x,n}^{+}e^{-\gamma_{\mathrm{f,n}}y}+A_{-,n}\Theta_{x,n}^{-}e^{+\gamma_{\mathrm{f,n}}(y-l)}, \tag{B5a}$$

$$M_y=\frac{E_0}{i\omega\mu_0}e^{ik_x x}\sum_{n=1}^{3}A_{+,n}\Theta_{y,n}^{+}e^{-\gamma_{\mathrm{f,n}}y}+A_{-,n}\Theta_{y,n}^{-}e^{+\gamma_{\mathrm{f,n}}(y-l)}, \tag{B5b}$$

with coefficients,

$$\theta_n^{\pm}=\frac{1}{\left(\mu_{t,n}^2-\mu_{g,n}^2\right)}\left(\mp\mu_{t,n}\gamma_{\mathrm{f,n}}+\mu_{g,n}k_x\right), \tag{B6a}$$

$$\Theta_{x,n}^{\pm}=\mp\gamma_{\mathrm{f,n}}-\frac{1}{\left(\mu_{t,n}^2-\mu_{g,n}^2\right)}\left(\mp\gamma_{\mathrm{f,n}}\mu_{t,n}+\mu_{g,n}k_x\right), \tag{B6b}$$

$$\Theta_{y,n}^{\pm}=-ik_x-\frac{1}{\left(\mu_{t,n}^2-\mu_{g,n}^2\right)}\left(\pm i\mu_{g,n}\gamma_{\mathrm{f,n}}-i\mu_{t,n}k_x\right). \tag{B6c}$$

Note that the magnetization vector is nonzero only in the ferrite region.

As usual, the dispersion relation of the guided modes is obtained by applying boundary conditions at the material interfaces. The standard Maxwellian boundary conditions for our system are i) the PEC boundary condition at the left wall ( $E_z|_{y=0}=0$ ), and ii) the continuity of $E_z$ and $H_x$ at the ferrite-air interface ( $y=l$ ). These yield the following equations:

$$\sum_{n=1}^{3} A_{+,n} + \sum_{n=1}^{3} A_{-,n} e^{-\gamma_{\mathrm{f,n}} l} = 0\,, \tag{B7a}$$

$$\sum_{n=1}^{3} A_{+,n} e^{-\gamma_{\mathrm{f,n}} l} + \sum_{n=1}^{3} A_{-,n} - A_0 = 0\,, \tag{B7b}$$

$$\sum_{n=1}^{3} A_{+,n} \theta_n^+ e^{-\gamma_{\mathrm{f,n}} l} + \sum_{n=1}^{3} A_{-,n} \theta_n^- - A_0 \gamma_0 \coth\left(\gamma_0 (l-a)\right) = 0\,, \tag{B7c}$$

In addition, under the nonlocal model, and in the absence of surface pinning or surface anisotropy, the magnetization vector is typically required to satisfy the Neumann boundary condition $\partial_y \mathbf{M} = 0$ at the two ferrite boundaries ( $y = 0, l$ ) [28]. This yields four additional boundary conditions, corresponding to the two transverse magnetization components at each interface ( $\partial_y M_x = \partial_y M_y = 0$ ):

$$\sum_{n=1}^{3} A_{+,n} \left(-\gamma_{\mathrm{f,n}} \Theta_{x,n}^+\right) + \sum_{n=1}^{3} A_{-,n} \gamma_{\mathrm{f,n}} \Theta_{x,n}^- e^{-\gamma_{\mathrm{f,n}} l} = 0\,, \tag{B7d}$$

$$\sum_{n=1}^{3} A_{+,n} \left(-\gamma_{\mathrm{f,n}} \Theta_{x,n}^+\right) e^{-\gamma_{\mathrm{f,n}} l} + \sum_{n=1}^{3} A_{-,n} \gamma_{\mathrm{f,n}} \Theta_{x,n}^- = 0\,, \tag{B7e}$$

$$\sum_{n=1}^{3} A_{+,n} \left(-\gamma_{\mathrm{f,n}} \Theta_{y,n}^+\right) + \sum_{n=1}^{3} A_{-,n} \gamma_{\mathrm{f,n}} \Theta_{y,n}^- e^{-\gamma_{\mathrm{f,n}} l} = 0\,, \tag{B7f}$$

$$\sum_{n=1}^{3} A_{+,n} \left(-\gamma_{\mathrm{f,n}} \Theta_{y,n}^+\right) e^{-\gamma_{\mathrm{f,n}} l} + \sum_{n=1}^{3} A_{-,n} \gamma_{\mathrm{f,n}} \Theta_{y,n}^- = 0\,. \tag{B7g}$$

The system of equations (B7) defines a set of seven homogeneous equations with seven unknowns. Setting the determinant of the corresponding coefficient matrix equal to zero yields the dispersion equation for the relevant waveguide modes.

## Disclosures

The authors declare no conflicts of interest.

## Data availability

All the data generated in this study are reported in the manuscript.